\documentclass[twocolumn,showpacs,amsmath,amssymb,pra]{revtex4}

\usepackage{amsxtra}
\usepackage{amssymb}
\usepackage{amsmath}
\usepackage{graphicx}
\usepackage[bf]{subfigure}

\newcommand {\rhovec}{\ensuremath \boldsymbol{\rho}}
\newcommand {\kvec}{\ensuremath \boldsymbol{k}}

\begin{document}
\title{Gaussian-state theory of two-photon imaging}
\date{\today}
\author{Baris I. Erkmen}
\email{erkmen@mit.edu}
\author{Jeffrey H. Shapiro}
\affiliation{Research Laboratory of Electronics, Massachusetts Institute of Technology, Cambridge, Massachusetts 02139, USA}

\begin{abstract}
Biphoton states of signal and idler fields---obtained from spontaneous parametric downconversion (SPDC) in the low-brightness, low-flux regime---have been utilized in several quantum imaging configurations to exceed the resolution performance of conventional imagers that employ coherent-state or thermal  light.   Recent work---using the full Gaussian-state description of SPDC---has shown that the same resolution performance seen in quantum optical coherence tomography and the same imaging characteristics found in quantum ghost imaging can be realized by classical-state imagers that make use of phase-sensitive cross correlations.  This paper extends the Gaussian-state analysis to two additional biphoton-state quantum imaging scenarios:  far field diffraction-pattern imaging; and broadband thin-lens imaging.  It is shown that the spatial resolution behavior in both cases is controlled by the nonzero phase-sensitive cross correlation between the signal and idler fields.  Thus, the same resolution can be achieved in these two configurations with classical-state signal and idler fields possessing a nonzero phase-sensitive cross correlation. 
\end{abstract}
\maketitle 

\section{Introduction}

Spontaneous parametric downconversion (SPDC), with a continuous-wave nondepleting pump, produces signal and idler fields that are in a maximally-entangled, zero-mean jointly Gaussian state \cite{Shapiro:Gaussian,Brambilla:SPDC}. When the downconverter is operated in its low-brightness, low-flux regime, so that at most one signal-idler photon pair is emitted during an observation interval, this state reduces to the superposition of a predominant multimode vacuum state plus a weak biphoton component \cite{Kim}.  Biphoton illumination is at the heart of several quantum imaging configurations, including quantum optical coherence tomography (OCT) \cite{Abouraddy}, quantum ghost imaging \cite{Pittman}, quantum holography \cite{Abouraddy:Holography} and quantum lithography \cite{Dangelo}.  These systems offer performance advantages over conventional optical imagers,  which employ coherent-state or thermal-state sources, that have traditionally been ascribed to the entanglement between the biphoton's signal and idler components.  We have shown---using Gaussian-state analysis and phase-sensitive coherence theory---that the advantages in quantum OCT and ghost imaging predominantly stem from the phase-sensitive cross correlation between the signal and idler fields, rather than their entanglement {\it per se} \cite{ErkmenShapiro:PCOCT,ErkmenShapiro:GhostImaging2}. Furthermore, because a pair of classical-state fields \cite{classicalstates}
can also have nonzero phase-sensitive cross correlation, most of the advantages seen in these biphoton-state imagers are also attainable with {\em classical} phase-sensitive sources, but conceivably (and conveniently) at much higher photon flux and without the need for single-photon counters.

In this paper we will study Fourier-plane and thin-lens imaging of a transmission mask using jointly-Gaussian source states.  These states encompass the biphoton state, thermal states (used in conventional low-coherence imaging) and coherent states (used in conventional coherent imaging). Previous theoretical and experimental work has shown that biphoton-state illumination of a transmission mask yields a far-field diffraction (Fourier-plane) pattern that is factor-of-two compressed relative to that produced by coherent plane-wave illumination of the mask \cite{Dangelo}. In addition, it has been claimed that imaging a transmission mask using broadband, spatially-incoherent biphoton-state illumination and a finite-diameter thin lens will yield a point-spread function that is a factor of two narrower than that obtained with quasimonochromatic spatially-incoherent thermal-state illumination \cite{Shih:QI}.  As noted above for quantum OCT and ghost imaging, these biphoton-state resolution enhancements in Fourier-plane and thin-lens imaging have been ascribed to signal-idler entanglement.  We will use Gaussian-state analysis, together with phase-sensitive coherence theory, to develop a unified---and generalized---understanding of the classical and quantum regimes of these imaging configurations. In particular, we will show that, once again, the benefits ascribed to entanglement are in fact due to phase-sensitive coherence at the source, and therefore are obtainable with classical-state sources  with phase-sensitive coherence. Furthermore, we will show that the narrowing of the point-spread function in the thin-lens imaging configuration is much less than a factor of two even at the theoretical upper  limit of SPDC bandwidth, and that it is marginal for typical SPDC bandwidths.  

The remainder of this paper is organized as follows. In Sec.~\ref{CH6:CoherenceTheory} we present the phase-insensitive and phase-sensitive coherence theory results that are relevant to Fourier-plane imaging.   In Sec.~\ref{CH6:FarFieldDiffraction} we use the Sec.~\ref{CH6:CoherenceTheory} results for quasimonchromatic light to determine the far-field diffraction properties of Gaussian-state source fields after they have illuminated a source-plane transmission mask. Section~\ref{CH6:BroadbandImaging} presents our analysis of thin-lens imaging, in which we determine the effect of the source's bandwidth on the point-spread functions resulting from phase-insensitive and phase-sensitive Gaussian-state illumination.  In doing so we pay particular attention to frequency-dependent propagation effects that did not enter into our quasimonochromatic treatment of Fourier-plane imaging.  Finally, in Sec.~\ref{CH6:discussion}, we summarize the results we have obtained for the Fourier-plane and thin-lens imaging configurations, highlighting the role played by phase-sensitive coherence. 

\section{Second-order coherence propagation}
\label{CH6:CoherenceTheory}

The imaging configurations we shall consider later acquire the far-field diffraction pattern or the transverse image of a transmission mask placed at the source's output plane. In both cases linear optical elements---a transmission mask, a lens, and a polarizing beam splitter---plus free-space propagation lie between the source and image-acquisition planes. Because the images are acquired via photocurrent correlation, their properties are determined by fourth-order correlations of the detected fields. To determine these correlation functions, it is necessary to propagate the fourth-order correlation function from the source plane to the appropriate image-acquisition planes. Fortunately,  zero-mean Gaussian source states are completely determined by their phase-insensitive and phase-sensitive, second-order auto- and cross-correlation functions. Therefore, we need only consider second-order coherence transfer.

Let $\hat{E}_{z}(\rhovec,t) e^{-i \omega_{0}t}$ denote a scalar, $z$-propagating, positive-frequency field operator with center-frequency $\omega_{0}$, and $\sqrt{\text{photons}/\text{m}^2 \text{s}}$ units. The commutators for the baseband envelope at a fixed transverse plane are given by \cite{Yuen:Part_I}
\begin{align}
[ \hat{E}_{z}(\rhovec_{1},t_{1}),\hat{E}_{z}(\rhovec_{2},t_{2})] & =0 
\label{commutator1}\\[5pt]
[ \hat{E}_{z}(\rhovec_{1},t_{1}),\hat{E}_{z}^{\dagger}(\rhovec_{2},t_{2})] & =\delta(\rhovec_{2} - \rhovec_{1}) \delta(t_{2} - t_{1})\,, \label{commutator2}
\end{align}
where $\delta(\cdot)$ is the impulse function.  

Free-space paraxial propagation is governed by the Huygens-Fresnel principle \cite{Yuen:Part_I}, which states that the baseband field operator at $z=L$ is related to the field operator at $z=0$ by the following superposition integral: 
\begin{eqnarray}
\hat{E}_{L}(\rhovec,t) &=& \int \! \frac{d\Omega}{2\pi}\, \int\! d \rhovec'\, e^{-i \Omega t} \hat{\mathcal{E}}_{0}\big(\rhovec',\Omega \big) \nonumber \\[.05in]  &\times& h_{L}(\rhovec-\rhovec',\omega_{0}+\Omega)  , \label{FS:prop}
\end{eqnarray}
where 
\begin{equation}
\hat{\mathcal{E}}_{z}(\rhovec,\Omega) \equiv \int \! dt \: \hat{E}_{z} (\rhovec,t) e^{i \Omega t} \label{FTR:FieldOperator}
\end{equation}
is the Fourier transform of the baseband-envelope field operator and
\begin{equation}
h_{L}(\rhovec,\omega) \equiv \frac{\omega}{i 2 \pi cL} e^{i \omega(L+ |\rhovec|^2/2L)/c}\,, \label{HFGreensFunction}
\end{equation}
is the Huygens-Fresnel Green's function for paraxial diffraction at frequency $\omega$ with $c$ being the vacuum light speed.  

For Eqs.~\eqref{FS:prop}--\eqref{HFGreensFunction} to be consistent with Eq.~\eqref{commutator2}, the integrals in Eqs.~\eqref{FS:prop} and \eqref{FTR:FieldOperator} must all be over infinite limits.  This poses no difficulty for the spatial integration, which can be taken over the entire $z=0$ plane, but there is a problem with the frequency integration.  Because $\hat{E}_z(\rhovec,t)$ is a positive-frequency field operator, the lower limit of integration in Eq.~\eqref{FS:prop} should be $-\omega_0$.  But, if we impose this propagation condition we do \em not\/\rm\ preserve the commutator in Eq.~\eqref{commutator2}, i.e., assuming that this commutator applies at $z=0$ and employing the diffraction integral does not recover the delta-function commutator at $z=L$.  
In almost all quantum optics situations---both theoretical and experimental---we can circumvent this issue as follows.  If the source at $z=0$ only excites frequencies that are within some bandwidth $\pm \Omega_0$ about $\omega_0$ and the measurements performed at $z=L$ are only sensitive to frequencies within that bandwidth, then so long as $\Omega_0 < \omega_0$, we can allow the frequency lower limit in Eq.~\eqref{FS:prop} to be $-\infty$.  In the quasimonochromatic cases to be considered below the preceding condition will be satisfied.  However, we will impose the finite lower limit on the frequency integral when we address thin-lens imaging at the ultimate theoretical limit of broadband SPDC.  In this case, the relevant commutators are
\begin{align}
[ \hat{\mathcal{E}}_{z}(\rhovec_{1},\Omega_{1}),\hat{\mathcal{E}}_{z}(\rhovec_{2},\Omega_{2})] & =0 
\label{OmegaCommutator1}\\[5pt]
[ \hat{\mathcal{E}}_{z}(\rhovec_{1},\Omega_1),\hat{\mathcal{E}}_{z}^{\dagger}(\rhovec_{2},\Omega_{2})] & =2\pi \delta(\rhovec_{2} - \rhovec_{1}) \delta(\Omega_{2} - \Omega_{1})\,. \label{OmegaCommutator2}
\end{align}

The non-Hermitian baseband field operator $\hat{E}_{z}(\rhovec,t)$ has two second-order correlation functions, namely the (normally-ordered) phase-insensitive correlation function
\begin{equation}
K^{(n)}_{z} (\rhovec_{1},\rhovec_{2}, t_{2}-t_{1}) \equiv \langle \hat{E}_{z}^{\dagger}(\rhovec_1,t_1) \hat{E}_{z}(\rhovec_2,t_2) \rangle\,, \label{PIScorr}
\end{equation}
and the phase-sensitive correlation function
\begin{equation}
K^{(p)}_{z} (\rhovec_{1},\rhovec_{2}, t_{2}-t_{1}) \equiv \langle \hat{E}_{z}(\rhovec_1,t_1) \hat{E}_{z}(\rhovec_2,t_2) \rangle\,, \label{PScorr}
\end{equation}
in which we have assumed, for simplicity, that the baseband field operator is in a complex-stationary state, viz., the correlation functions depend on the time difference $t_{2}-t_{1}$, but not on the absolute times.

We will find it convenient and insightful to work with the frequency spectra associated with Eqs.~\eqref{PIScorr} and \eqref{PScorr}, defined as the Fourier transforms,
\begin{equation}
S_{z}^{(x)} (\rhovec_{1}, \rhovec_{2}, \Omega) \equiv \int\!d\tau \: K_{z}^{(x)}(\rhovec_{1}, \rhovec_{2}, \tau) e^{i \Omega \tau}\,, \label{CH6:Spectra}
\end{equation}
for $x = n,p $. The phase-insensitive and phase-sensitive correlation spectra at $z=L$ can be expressed in terms of the correlation spectra at $z=0$ by evaluating Eq.~\eqref{CH6:Spectra} for the propagated field operators, via Eq.~\eqref{FS:prop}, which yields the following phase-insensitive spectrum,
\begin{eqnarray}
\lefteqn{S_{L}^{(n)} (\rhovec_{1}, \rhovec_{2}, \Omega) = \iint d\rhovec_{1}'\, d\rhovec_{2}' \, S_{0}^{(n)} (\rhovec_{1}', \rhovec_{2}', \Omega)} \nonumber \\[5pt] &\times& h_{L}^{*}(\rhovec_{1}-\rhovec_{1}',\omega_{0} + \Omega) h_{L}(\rhovec_{2}-\rhovec_{2}',\omega_{0} + \Omega)\,, \label{PIS:spectrum}
\end{eqnarray}
and the following phase-sensitive spectrum,
\begin{eqnarray}
\lefteqn{S_{L}^{(p)} (\rhovec_{1}, \rhovec_{2}, \Omega) = \iint d\rhovec_{1}'\, d\rhovec_{2}' \, S_{0}^{(p)} (\rhovec_{1}', \rhovec_{2}', \Omega)} \nonumber \\[5pt] &\times& h_{L}(\rhovec_{1}-\rhovec_{1}',\omega_{0} - \Omega) h_{L}(\rhovec_{2}-\rhovec_{2}',\omega_{0} + \Omega)\,, \label{PS:spectrum}
\end{eqnarray}
where $*$ in Eq.~\eqref{PIS:spectrum} denotes complex conjugation. Note that the phase-insensitive correlation spectrum is a monochromatic equation, i.e., the frequency dependence is $\omega_{0} + \Omega$ on both sides of the equality, whereas the phase-sensitive spectrum is a bichromatic equation involving $\omega_{0} \pm \Omega$. This difference occurs because complex-stationary phase-insensitive correlations in time have uncorrelated frequency components, whereas complex-stationary phase-sensitive correlation functions in time have nonzero (phase-sensitive) correlations between frequency components with equal and opposite detunings from the field's center frequency \cite{ErkmenShapiro:PSCoherenceThy}. 

Consider a quasimonochromatic state in which the preceding spectra are only nonzero for $| \Omega | /\omega_{0} \ll 1$. It follows that the Huygens-Fresnel principle simplifies to
\begin{equation}
\hat{E}_{L}(\rhovec,t) \!=\! \int\! d\rhovec'\, \hat{E}_{0}\big(\rhovec',t-L/c \big) h_{L}(\rhovec-\rhovec',\omega_{0}) , \label{FS:propQM}
\end{equation}
and the propagation Green's function in Eqs.~\eqref{PIS:spectrum} and \eqref{PS:spectrum} become  $h_{L}(\rhovec, \omega_{0} \pm \Omega) \approx h_{L}(\rhovec,\omega_{0})$. 
Let us now review the far-field propagation regime for this quasimonochromatic situation.  We will 
assume that the field at the $z=0$ plane is in a zero-mean state with cross-spectrally pure, Schell-model  correlation spectra \cite{coherence_separability} given by
\begin{align}
S_{0}^{(n)}(\rhovec_1, \rhovec_2,\Omega) &\!=\! T^{*}\!(\rhovec_1) T(\rhovec_2) G^{(n)}(\rhovec_2\!-\!\rhovec_1) S^{(n)} (\Omega),  \label{SM0:PIS} \\[5pt]
S_{0}^{(p)}(\rhovec_1, \rhovec_2,\Omega)  & \!=\!  T(\rhovec_1) T(\rhovec_2) G^{(p)}(\rhovec_2\!-\!\rhovec_1) S^{(p)}(\Omega).\!\label{SM0:PS}
\end{align}
With no loss of generality, we require $|T(\rhovec)| \leq 1$, so that it may be regarded as a (possibly complex-valued) spatial attenuation of an optical field operator in a homogenous and stationary state with separable phase-insensitive spectrum $S^{(n)}(\Omega) G^{(n)}(\rhovec_{2}-\rhovec_{1})$ and phase-sensitive spectrum $S^{(p)}(\Omega) G^{(p)}(\rhovec_{2}-\rhovec_{1})$.  This spatial attenuation  will become the transmission mask to be imaged when we turn our attention to the Fourier-plane and thin-lens imaging configurations.  For now, however, it is convenient to lump this mask together with the source.  

Our primary interest is in sources with narrow $G^{(x)}_{0}(\rhovec)$, for $x=n,p$, such that $T(\rhovec)$ does not vary appreciably within a (phase-insensitive or phase-sensitive) coherence area.
For this case, we may approximate the source correlation spectra as follows:  
\begin{align}
S_{0}^{(n)}(\rhovec_1, \rhovec_2,\Omega)  & \approx \bigl | T(\rhovec_{s} )\bigr |^2 G^{(n)}(\rhovec_d) S^{(n)}(\Omega),   \label{SM02:PIS}\\[5pt]
S_{0}^{(p)}(\rhovec_1, \rhovec_2,\Omega) &\approx T^{2} ( \rhovec_s  ) G^{(p)}(\rhovec_d) S^{(p)}(\Omega) \,,\label{SM02:PS}
\end{align}
in terms of the sum coordinate $\rhovec_{s} \equiv (\rhovec_{2} + \rhovec_{1}) /2 $ and the difference coordinate $\rhovec_{d} \equiv \rhovec_{2} - \rhovec_{1}$. This approximation simplifies the subsequent analytic treatment considerably, without significant impact on the fundamental physics. The $z=L$ spectra then become
\begin{align}
\lefteqn{S_{L}^{(n)}(\rhovec_{1}, \rhovec_{2}, \Omega) = \frac{\omega_{0}^{2} S^{(n)}(\Omega) }{(2 \pi c L)^2}  \, e^{i  \omega_{0} \rhovec_{s}\cdot \rhovec_{d}/cL}} \nonumber \\[5pt] &&\times \int \!d\rhovec_{s}' \int\! d\rhovec_{d}' \, e^{-i  \omega_{0} (\rhovec_{s}\cdot \rhovec_{d}' + \rhovec_{d} \cdot \rhovec_{s}')/cL}\,  e^{i \omega_{0} \rhovec_{s}' \cdot \rhovec_{d}'/cL} \nonumber \\[5pt] && \times |T(\rhovec_{s}')|^2 G^{(n)}(\rhovec_{d}'), \label{PIS:CoherencePropagation}
\end{align}
and
\begin{align}
S_{L}^{(p)} (\rhovec_{1}, &\rhovec_{2}, \Omega) =  \frac{-\omega_{0}^{2} S^{(p)}(\Omega)}{(2 \pi cL)^2} \,  e^{i  \omega_{0} \left (2 L^{2} + |\rhovec_{s}|^2+ |\rhovec_{d}|^2/4 \right )/cL}  \nonumber \\[5pt] &\times \int \!d\rhovec_{s}' \int \!d\rhovec_{d}'  \, e^{-i \omega_{0} (2 \rhovec_{s}\cdot \rhovec_{s}' +  \rhovec_{d} \cdot \rhovec_{d}'/2)/cL}  \nonumber \\[5pt] & \times  e^{i  \omega_{0} (|\rhovec_{s}'|^2+ |\rhovec_{d}'|^2/4)/cL}  \, T^2(\rhovec_{s}') G^{(p)}(\rhovec_{d}'), \label{PS:CoherencePropagation}
\end{align}
respectively, where the quasimonochromatic assumption has permitted our approximating the frequency-dependent leading coefficients by their values at the center frequency \cite{ApproxValidation}. 

Let $\rho_{0}$ denote the coherence radius of the source, which we shall assume to be the same for both phase-insensitive and phase-sensitive correlations, i.e., $\rho_0$ is the radius within which $G^{(x)}(\rhovec)$, for $x=n,p$, differ appreciably from zero.   Also, let $a_{0} \gg \rho_0$ denote the transverse radius of $|T(\rhovec)|^{2}$, which will be the photon-flux density radius of the source's state just after the transmission mask in our imaging configurations.   In far field phase-insensitive correlation propagation, which applies when $\omega_{0} a_{0} \rho_{0}/2cL \ll 1$, the phase term $e^{i \omega_{0} \rhovec_{s}' \cdot \rhovec_{d}'/cL}$ can be neglected in Eq.~\eqref{PIS:CoherencePropagation}, and we find that 
\begin{eqnarray}
S_{L}^{(n)} (\rhovec_{1}, \rhovec_{2}, \Omega) &=& \frac{\omega_{0}^{2} S^{(n)}(\Omega)}{(2 \pi cL)^2} e^{i \omega_{0} \rhovec_{s}\cdot \rhovec_{d}/cL} \nonumber \\[5pt] &\times&  \mathcal{T}_{n}\!\left (\frac{\omega_{0} \rhovec_{d}}{cL} \right ) \mathcal{G}^{(n)}\!\left (\frac{\omega_{0} \rhovec_{s}}{cL} \right )\,,  \label{VCZ:PIS}
\end{eqnarray}
where $\mathcal{T}_{n}(\kvec)$ and $\mathcal{G}^{(n)}(\kvec)$ are the $2$-D Fourier transforms,
\begin{align}
\mathcal{T}_{n}(\kvec) & \equiv \int\!{d}\rhovec' e^{-i \kvec \cdot \rhovec'} |T(\rhovec')|^{2}\,, \label{Tn}\\[5pt]
\mathcal{G}^{(n)} (\kvec) &\equiv \int\! {d}\rhovec' e^{-i \kvec \cdot \rhovec'} G^{(n)} (\rhovec')\,.
\end{align}
The Fourier-transform duality between the source-plane and the far-field phase-insensitive correlation spectra---seen in Eq.~\eqref{VCZ:PIS}---is the well known van Cittert-Zernike theorem for phase-insensitive correlation propagation \cite{Mandel}.  A similar duality is present between the source-plane and the far-field phase-sensitive correlation spectra, but the far-field regime---in which the quadratic phase terms of the integrand in Eq.~\eqref{PS:CoherencePropagation} become negligible---corresponds to $\omega_{0} a_{0}^{2}/2cL  \ll 1$, which is more stringent than the far-field condition for the phase-insensitive case. In this regime, we find that
\begin{eqnarray}
S_{L}^{(p)} (\rhovec_{1}, \rhovec_{2}, \Omega) &=& \frac{-\omega_{0}^{2} S^{(p)}(\Omega)}{(2 \pi cL)^2} \,e^{i \omega_{0} \left ( 2 L^{2} + |\rhovec_{s}|^2+ |\rhovec_{d}|^2/4 \right)/cL} \nonumber \\[5pt] &\times&\mathcal{T}_{p}\left (\frac{2 \omega_{0}  \rhovec_{s} }{cL} \right ) \mathcal{G}^{(p)}\! \left ( \frac{\omega_{0} \rhovec_{d}}{2cL} \right ), \label{VCZ:PS}
\end{eqnarray}
gives the far-field phase-sensitive correlation spectrum, with
\begin{equation}
\mathcal{T}_{p}(\kvec)  \equiv \int\!{d}\rhovec' e^{-i \kvec \cdot \rhovec'} T^{2}(\rhovec') 
\end{equation}
and 
\begin{equation}
\mathcal{G}^{(p)} (\kvec) \equiv \int \!{d}\rhovec' e^{-i \kvec \cdot \rhovec'} G^{(p)} (\rhovec')\,.
\end{equation}
By analogy  to the phase-insensitive case, we refer to the Fourier transform relation in Eq.~\eqref{VCZ:PS} as the van Cittert-Zernike theorem for phase-sensitive coherence propagation. 

To conclude our review of far-field coherence propagation it is worth emphasizing the similarities and differences  between Eqs.~\eqref{VCZ:PIS} and \eqref{VCZ:PS}. The source-plane transmission mask, $T(\rhovec)$, has been assumed to be a slowly-varying and broad function in comparison to the rapidly decaying $G^{(x)}(\rhovec)$. Thus, for $x = n, p$, Fourier-transform duality implies that $\mathcal{T}_{x}(\kvec)$, decays more rapidly than $\mathcal{G}^{(x)}(\kvec)$. Therefore, the far-field phase-sensitive correlation function consists of a narrow function of $\rhovec_{s}$ multiplying a broad function of $\rhovec_{d}$, whereas the corresponding phase-insensitive correlation function consists of a narrow function of $\rhovec_{d}$ times a broad function of $\rhovec_{s}$. Owing to this difference, point pairs on the transverse plane in the far field that are symmetrically disposed about the origin, viz., points satisfying $|\rhovec_{s}| \approx 0$, have appreciable phase-sensitive correlation.  The phase-insensitive correlation, however, is highest between point pairs that are in close proximity on the transverse plane, i.e., point pairs obeying $|\rhovec_{d}| \approx 0$.  In addition, if we evaluate the correlations at a single transverse point, i.e., when $\rhovec_{d} = {\bf 0}$, we find that the phase-insensitive correlation traces out the broad envelope $\mathcal{G}^{(n)}(\omega_0\rhovec_s/cL)$, whereas the phase-sensitive correlation traces out the narrow function $\mathcal{T}_{p}(2\omega_0\rhovec_s/cL)$, a property we shall make use of in the following section. Finally, it is relevant to emphasize that Eq.~\eqref{VCZ:PS} is a general property of phase-sensitive coherence propagation that applies regardless of whether the source state is classical or nonclassical  \cite{ErkmenShapiro:PSCoherenceThy,ErkmenShapiro:GhostImaging2}.

\section{Two-Photon Diffraction Pattern Imaging} 
\label{CH6:FarFieldDiffraction}

\begin{figure}[t]
\begin{center}
\includegraphics[width= 3.0in]{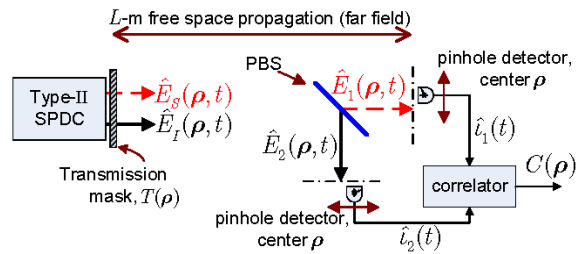}
\end{center}
\caption{(Color online) Imaging the far-field diffraction pattern of a transmission mask.  PBS, polarizing beam splitter.} \label{TPI:FFdiff} 
\end{figure}

Consider the experimental setup, shown in Fig.~\ref{TPI:FFdiff}, whose purpose is to obtain the far-field diffraction pattern of a source-plane transmission mask of (possibly complex-valued) field transmissivity $T(\rhovec)$. An experiment using this setup---which we will explain in detail shortly---was reported in \cite{Dangelo} as a proof-of-principle demonstration of biphoton-state quantum lithography.  That experiment exhibited a factor-of-two compression in the fringe pattern produced using a two-slit transmission mask as compared to what was obtained with conventional (coherent-state) illumination at the same wavelength of that two-slit mask.  The observed fringe-pattern compression was therefore interpreted as the expected result for quantum lithography using a pair of entangled photons \cite{QuantLithFootnote}.  Our aim in this section is to show that it is phase-sensitive coherence, not entanglement {\it per se}, that is responsible for this fringe-pattern compression. In particular, we will find that classical phase-sensitive light and biphoton-state light yield \em identical\/\rm\ images, except for the image produced with the classical-state source being embedded in a prominent featureless background that is absent for the case of biphoton illumination. Thus, the fringe-pattern compression previously ascribed to the entangled nature of the signal and idler photons is in fact due to their having a phase-sensitive cross correlation, which is a property that classical states can also possess. Signal-idler entanglement, which in the Gaussian-state framework is a stronger-than-classical phase-sensitive cross correlation between the signal and idler fields, is responsible for dramatically improving the contrast of this image when the illumination is broadband.

Now let us flesh out the details of the preceding assertions within the context of the Fig.~\ref{TPI:FFdiff} setup.  The source will be taken to be frequency-degenerate type-II phase-matched SPDC with a continuous-wave pump.  It generates paraxial, $z$-propagating signal ($S$) and idler ($I$) fields in orthogonal polarizations with common center frequency $\omega_{0}$. Their positive-frequency field operators will be denoted as $\hat{E}_{S}(\rhovec,t)e^{-i \omega_{0}t}$ and $\hat{E}_{I}(\rhovec,t)e^{-i \omega_{0}t}$. With a nondepleting plane-wave pump and ignoring boundary effects due to the nonlinear crystal's finite cross-section, these two output fields are in a zero-mean, jointly Gaussian state that is homogeneous and stationary.  Moreover, the signal and idler will then have identical fluorescence spectra and maximum phase-sensitive cross correlation, but no phase-sensitive autocorrelation or phase-insensitive cross correlation \cite{Shapiro:Gaussian,Brambilla:SPDC}. It follows that after passing through the transmission mask the jointly Gaussian state is fully determined by the Schell-model autocorrelation spectra \cite{coherence_separability}
\begin{equation}
S_{m,m}^{(n)}(\rhovec_1, \rhovec_2,\Omega)= S_{0}^{(n)}(\rhovec_1, \rhovec_2,\Omega)\,, \label{PIS:Source}
\end{equation}
for $m = S, I$,  and the phase-sensitive cross-correlation spectrum 
\begin{equation}
S_{S,I}^{(p)}(\rhovec_1, \rhovec_2,\Omega) = S_{0}^{(p)}(\rhovec_1, \rhovec_2,\Omega)\,, \label{PS:Source}
\end{equation}
where $S_{0}^{(n)}$ and $S_{0}^{(p)}$ are given  by Eqs.~\eqref{SM02:PIS} and \eqref{SM02:PS} respectively, and we shall assume that they satisfy the quasimonochromatic condition.  

In the Fig.~\ref{TPI:FFdiff} setup, the signal and idler fields both propagate over an $L\,$m free-space path---assumed to be sufficiently long to satisfy the far-field condition, $\omega_0a_0^2/2cL \ll 1$, for the phase-sensitive source---and then are separated by a polarizing beam splitter such that they impinge on separate pinhole detectors, each centered on the transverse-plane coordinate $\rhovec$ with respect to its optical axis.  Because linear transformations of zero-mean Gaussian states are still zero-mean Gaussian \cite{Shapiro:Gaussian}, and because free-space diffraction is a linear transformation, we need only determine the second-order moments at the detection planes to determine the joint state of $\hat{E}_{1}(\rhovec,t)$ and $\hat{E}_{2}(\rhovec,t)$, which denote the far-field propagated field operators of the signal and the idler respectively. Thus, quasimonochromatic paraxial diffraction into the far field results in phase-insensitive autocorrelation spectra given by Eq.~\eqref{VCZ:PIS} and a phase-sensitive cross-correlation spectrum given by Eq.~\eqref{VCZ:PS} \cite{AutovsCross}.

The two pinhole photodetectors are assumed to have identical parameters: quantum efficiency $\eta$; photosensitive area $A$; and current-pulse output $qh_B(t)$ from detection of a single photon, where $q$ is the electron charge and $\int\!dt\,h_B(t) = 1$. Then, the time-average photocurrent cross-correlation at the detection planes has an ensemble average \cite{ShapiroSun,ErkmenShapiro:GhostImaging2}
	\begin{equation}
 C(\rhovec) = \frac{1}{T} \int_{-T/2}^{T/2} {d}t \, \langle \hat{\imath}_{1}(t) \hat{\imath}_{2}(t) \rangle\,, \label{coinc}
\end{equation}
in which the equal-time photocurrent cross correlation is given by
\begin{multline}
\langle \hat{\imath}_{1}(t) \hat{\imath}_{2}(t) \rangle = q^2 \eta^{2} A^{2} \int\! \int \! du_{1}  du_{2} \\[5pt] \times \langle \hat{E}_{1}^{\dagger}(\rhovec,u_{1}) \hat{E}_{2}^{\dagger}(\rhovec,u_{2}) \hat{E}_{1}(\rhovec,u_{1}) \hat{E}_{2}(\rhovec,u_{2})  \rangle \\[5pt] \times  h_{B}(t-u_{1}) h_{B}(t-u_{2})\,. \label{IntCorr}
\end{multline}
Here, we have approximated the integrals over the pinhole detectors' photosensitive regions by the value of the integrand at $\rhovec$ times $A^{2}$.  For our Gaussian-state $\hat{E}_{1}$ and $\hat{E}_{2}$, the fourth-order field moment in Eq.~\eqref{IntCorr} reduces to a sum of products of second-order correlation functions by virtue of the Gaussian moment-factoring theorem \cite{Mandel,ShapiroSun}. This procedure simplifies the photocurrent cross-correlation expression to 
\begin{equation}
C(\rhovec) = C_{0}(\rhovec) + C_{p} \left | \mathcal{G}^{(p)} ( \mathbf{0}  ) \: \mathcal{T}_{p}\left (\frac{2 \omega_{0}  \rhovec }{cL} \right ) \right | ^{2}\,. \label{TPI:corr}
\end{equation}
Here, 
\begin{eqnarray}
C_{0}(\rhovec) &=& \biggl [ \frac{\omega_{0}^{2} q \eta A}{4 \pi^2 c^2L^2} \, \mathcal{T}_{n}( \mathbf{0} )\, \mathcal{G}^{(n)}\! \left (\frac{\omega_{0} \rhovec}{cL} \right ) \nonumber \\[5pt] &\times&  \int_{-\omega_{0}}^{\infty} \! \frac{d\Omega}{2\pi} \,  S^{(n)}(\Omega)\, \int_{-\infty}^{\infty} \! dt \, h_{B}(t)  \biggr ]^{2}  \label{Co}
\end{eqnarray}
is a non-image-bearing background, which is broad and featureless owing to ${\mathcal{G}}^{(n)}({\boldsymbol k})$ being the Fourier transform of the narrow spatial-domain correlation function $G^{(n)}(\rhovec)$.  
The second term in Eq.~\eqref{TPI:corr} is the image-bearing term.   Its constant factor is 
\begin{equation}
C_{p} = \left ( \frac{\omega_{0}^{2} q \eta A}{4 \pi^{2} c^2L^{2}}\right )^{\!2} \! \left [ \bigl \lvert \mathcal{F}^{-1} \{ S^{(p)}(\Omega)\}  \bigr \rvert ^{2}  \!\star h_{B} \star \overleftarrow{h}_{\!\!B} \right ]_{t=0},\label{Cp}
\end{equation}
where $\mathcal{F}^{-1}\{ \cdot \}$ denotes the inverse Fourier transform of the bracketed term [see Eq.~\eqref{CH6:Spectra} for our Fourier transform sign convention], $\star$ denotes convolution, and $\overleftarrow{h}_{\!\!B}$ represents the time-reversed impulse response. From Eq.~\eqref{TPI:corr}, we see that the image-bearing term is proportional to $| \mathcal{T}_{p}(2 \omega_{0} \rhovec / cL)|^{2}$, which is the far-field diffraction pattern of the {\em square} of the mask's field transmissivity $T(\rhovec)$.

Let us compare the imaging characteristics of this imager to those of a conventional classical imager that utilizes a coherent-state beam to illuminate the mask, and a single (scanning) pinhole detector located in the far field that records the diffraction pattern. If we assume the field impinging on the transmission mask is a monochromatic plane wave at center frequency $\omega_{0}$ and with photon-flux density $I_{0}$, the field just after the mask is in the coherent state satisfying
\begin{equation}
\hat{E}_{0}(\rhovec, t) \lvert \sqrt{I_{0}} T(\rhovec)\rangle  = \sqrt{I_{0}} T(\rhovec) \lvert \sqrt{I_{0}} T(\rhovec)\rangle\,.
\end{equation}
Because free-space propagation is a multimode beam splitter relation \cite{Yuen:Part_I}, the detection-plane field operator $\hat{E}_{1}(\rhovec,t)$ is also in a coherent state, whose eigenfunction is determined by substituting $\sqrt{I_{0}} T(\rhovec)$ into the classical Huygens-Fresnel diffraction integral, i.e., Eq.~\eqref{FS:prop} with the field operator replaced by the coherent-state eigenfunction.

We shall assume the path length $L$ satisfies the far-field condition, $\omega_{0} a_{0}^{2}/2 cL  \ll 1$, for coherent-state diffraction \cite{NearvsFar:Coherent}, so that the quadratic phase term in the Huygens-Fresnel Green's function becomes negligible and the mean photocurrent becomes 
\begin{equation}
\langle \hat{\imath}(t) \rangle = \frac{\omega_{0}^{2} q \eta A}{4 \pi^{2} c^2L^{2}} \, I_{0} \int_{-\infty}^{\infty} dt \, h_{\!B}(t) \, \left | \mathcal{T}_{c}\left (\frac{\omega_{0} \rhovec}{cL} \right ) \right |^{2}\,,
\end{equation}
where there is no background and the image term is given by, 
\begin{equation}
\mathcal{T}_{c}(\kvec) \equiv \int_{\mathbb{R}^{2}} d\rhovec' \, e^{-i \kvec \cdot \rhovec'} T(\rhovec')\,.
\end{equation}
If $T(\rhovec)$ only takes values zero or one---as was the case for the two-slit transmission mask employed in \cite{Dangelo}---then $T^{2}(\rhovec) = T(\rhovec)$ and the biphoton source yields a far-field diffraction pattern proportional to $| \mathcal{T}_{p}(2 \omega_{0} \rhovec / cL)|^{2}$, whereas the coherent diffraction pattern is proportional to $| \mathcal{T}_{p}(\omega_{0} \rhovec / cL)|^{2}$. Thus, the far-field pattern observed with the biphoton source is spatially compressed by a factor of two relative to that obtained when the coherent-state source is employed, which is why the biphoton case has been said to beat the classical resolution limit. However, it is worth re-emphasizing that Eq.~\eqref{TPI:corr} is true for both classical and quantum Gaussian-state sources whose signal and idler fields have a nonzero phase-sensitive cross correlation.  Consequently, it is the phase-sensitive coherence of the source---and {\em not} the entanglement of the biphoton---that is the fundamental cause for the factor-of-two compression in the far-field diffraction pattern as compared to what is obtained using a coherent-state plane wave.  Finally, it is important to note that unless $T^2(\rhovec) \propto T(\rhovec)$, as we have above when $T(\rhovec)$ is a zero-one function, then the diffraction pattern acquired from phase-sensitive sources will be distorted relative to what will be obtained with a coherent-state field.

Utilizing phase-insensitive Gaussian-state light in the Fig.~\ref{TPI:FFdiff} configuration does not result in a photocurrent cross correlation containing a diffraction-pattern image.  This imaging failure occurs because, from Eq.~\eqref{VCZ:PIS}, the equal-position correlation in the Fourier plane traces out $\mathcal{G}_{0}^{(n)}(\omega_0 \rhovec/cL)$, which does not contain any information about the transmission mask $T(\rhovec)$.  However, by modifying the Fig.~\ref{TPI:FFdiff} setup to have one detector scan $-\rhovec$, while the other scans $\rhovec$, then the photocurrent cross correlation will contain an image of  $| \mathcal{T}_{n}(2 \omega_{0} \rhovec / cL )|^{2}$.  For a zero-one function transmission mask, this phase-insensitive imager also achieves the factor-of-two pattern compression in comparison with the image formed using a coherent-state plane wave.  In this regard we note that \em \/\rm\ phase-insensitive Gaussian-state sources are classical \cite{ErkmenShapiro:GhostImaging2}.

Thus far we have considered only the image-bearing term in Eq.~\eqref{TPI:corr}. Now we will address the image contrast. For simplicity, we will assume that $T(\rhovec)$  is real valued.  In addition, we restrict ourselves to an observation region ${\cal{R}}$ that encompasses the image-bearing term in Eq.~\eqref{TPI:corr}, and we define the contrast as
\begin{equation}
{\cal{C}} \equiv  \frac{\max_{\cal{R}}[C(\rhovec)] - \min_{\cal{R}}[C(\rhovec)]}{C_0({\bf 0})}\,,
\end{equation}
so that the numerator yields the dynamic range of the image-bearing term in the photocurrent correlation $C(\rhovec)$, while the denominator is the featureless background.

Here we compare the contrast from classical and quantum sources that have identical autocorrelation spectra and the maximum phase-sensitive cross correlation allowed in classical and quantum physics, respectively. When the source is in a classical Gaussian state, whose autocorrelation spectrum---just before the transmission mask---is $\mathcal{G}^{(n)}(\kvec) S^{(n)}(\Omega)$, the maximum magnitude for the phase-sensitive cross-correlation spectrum is equal to the autocorrelation spectrum \cite{Shapiro:Gaussian}, i.e., the Gaussian state with maximum classical phase-sensitive cross correlation satisfies
\begin{equation}
|S^{(p)}(\Omega) \mathcal{G}^{(p)}(\kvec)| = S^{(n)}(\Omega) \mathcal{G}^{(n)}(\kvec). \label{PS:classical}
\end{equation}
Taking the phase of this phase-sensitive spectrum to be zero, and recalling that $\int dt\, h_{B} (t) = 1$, the contrast with classical phase-sensitive Gaussian-state sources can be written in the form
\begin{equation}
{\cal{C}}^{(c)} = {\cal{C}}_s^{(c)}{\cal{C}}_t^{(c)}, 
\end{equation}
where the spatial ($s$) factor is given by
\begin{equation}
\mathcal{C}^{(c)}_{s} = \frac{\max_{\kvec}[ |\mathcal{T}_{p}(\kvec)|^{2} ] - \min_{\kvec}[ |\mathcal{T}_{p}(\kvec)|^{2} ]}{\mathcal{T}_{n}^{2}(\mathbf{0})} \leq 1\,, 
\end{equation}
with equality if $T(\rhovec)$ is real, so that
\begin{equation}
\mathcal{C}^{(c)} = \mathcal{C}_{t}^{(c)} = \frac{ \left [ \bigl \lvert \mathcal{F}^{-1} \{ S^{(n)}(\Omega)\}  \bigr \rvert ^{2}\!\star h_{B} \star \overleftarrow{h}_{\!\!B} \right ]_{t=0}}{\Bigl ( \int\! d\Omega\,  S^{(n)}(\Omega)/2\pi \Bigr )^{2}}
\end{equation}
for such masks.
For analytical convenience, let us take the spectral part of the phase-insensitive autocorrelation function to be Gaussian with $e^{-2}$-attenuation baseband bandwidth $2/T_{0}$, i.e., 
\begin{equation}
S^{(n)}(\Omega) =  e^{-T_{0}^{2} \Omega^{2}/2}\,\sqrt{2 \pi T_{0}^{2}}\,, \label{Sn:Gaussian}
\end{equation}
and let us take the baseband impulse response $h_B(t)$ to be a Gaussian with $e^{-2}$-attenuation time duration $T_d$, viz., 
\begin{equation}
h_B(t) = e^{-8t^2/T_d^2} \sqrt{8/\pi T_d^2} \,.  \label{Hb:Gaussian}
\end{equation}
With these assumptions, we find that the classical contrast for a real-valued mask is
\begin{equation}
\mathcal{C}^{(c)}  = \frac{1} { \sqrt{1 + (T_{d}/2 T_{0})^{2}}}\,,
\end{equation}
which is approximately unity for narrowband sources that satisfy $T_{d} \ll T_{0}$. On the other hand, in the broadband limit $T_{d} \gg T_{0}$, we have 
\begin{equation}
\mathcal{C}^{(c)} \approx 2  T_{0} / T_{d} \ll 1\,,
\end{equation}
so the contrast is severely degraded in this case.

Now consider a nonclassical Gaussian state with the maximum phase-sensitive cross correlation. In the low-brightness regime, i.e., when  $S^{(n)}(\Omega) \mathcal{G}_{0}^{(n)}(\kvec) \ll 1$, the maximum phase-sensitive cross-correlation spectrum is approximately \cite{ErkmenShapiro:GhostImaging2}
\begin{equation}
|S^{(p)}(\Omega) \mathcal{G}^{(p)}(\kvec)| \approx \sqrt{S^{(n)}(\Omega) \mathcal{G}^{(n)}(\kvec)},
\end{equation}
which is much higher, in this limit, than the classical maximum given by Eq.~\eqref{PS:classical}. Taking the phase of this correlation to be zero, the contrast is found to factor into the product of spatial and temporal terms, with $\mathcal{C}^{(q)}_{s} =\mathcal{C}^{(c)}_{s}$, and the temporal term given by
\begin{equation}
\mathcal{C}^{(q)}_{t} = \frac{ \left [ \bigl \lvert \mathcal{F}^{-1} \{ \sqrt{S^{(n)}(\Omega)} \}  \bigr \rvert ^{2}\!\star h_{B} \star \overleftarrow{h}_{\!\!B} \right ]_{t=0}}{\mathcal{G}^{(n)}(\mathbf{0}) \left (  \int\! d\Omega\,  S^{(n)}(\Omega)/2\pi \right )^{2}}\,.
\end{equation}
Once again using Eq.~\eqref{Sn:Gaussian} for the fluorescence spectrum and Eq.~\eqref{Hb:Gaussian} for the baseband current filter, we obtain
\begin{equation}
\mathcal{C}^{(q)} = 2 / \mathcal{G}^{(n)}(\mathbf{0}) S^{(n)}(0)   \sqrt{1 + T_{d}^{2} / 2 T_{0}^{2}}\,
\end{equation}
for real-valued $T(\rhovec)$.  
Here, the narrowband contrast $\mathcal{C}^{(q)} = 2/ \mathcal{G}^{(n)}(\mathbf{0}) S^{(n)}(0)$ is very high because of the low-brightness condition, and even for broadband fields the contrast,
\begin{equation}
\mathcal{C}^{(q)} = 2\sqrt{2} T_{0} / T_{d} \mathcal{G}^{(n)}(\mathbf{0}) S^{(n)}(0) \,,
\end{equation}
may be high. In particular, in the biphoton regime, wherein $\mathcal{G}^{(n)}(\mathbf{0}) T_{d} \ll 1$ (low flux as well as low brightness) also prevails, very high contrast is predicted in this broadband limit \cite{LowFluxCond}, which is in agreement with the background-free diffraction pattern reported in \cite{Dangelo}. Therefore, low-brightness quantum Gaussian-state fields have a contrast advantage over classical phase-sensitive Gaussian-state fields when the phase-sensitive cross correlation is measured via a photocurrent correlation measurement, and the biphoton state yields images with negligible background even when it is a broadband state.

In summary, in this section we have studied a cornerstone proof-of-principle experiment for quantum optical lithography by applying Gaussian-state analysis and the coherence theory results from Sec.~\ref{CH6:CoherenceTheory} to the propagation of classical and quantum phase-sensitive cross correlations. Our analysis has shown that the {\em only} performance difference between using a biphoton-state source and a classical phase-sensitive Gaussian-state source in the Fig.~1 setup is the diffraction-pattern image's contrast. The resolution improvement seen with a biphoton source is entirely due to the diffraction properties of the phase-sensitive cross correlation between the signal and idler fields, hence it is also achievable with a pair of classical Gaussian-state fields with phase-sensitive cross correlation. However, low-brightness quantum sources achieve higher contrast than classical sources, which permits imaging with broader bandwidth quantum sources. Finally, the broadband biphoton state yields very high contrast images, which is the reason why biphoton-state quantum lithography experiments have yielded background-free diffraction-pattern images~\cite{Dangelo}.

\section{Two-Photon Thin-Lens Imaging}
\label{CH6:BroadbandImaging}

\begin{figure}[t]
\begin{center}
\includegraphics[width= 3.0 in]{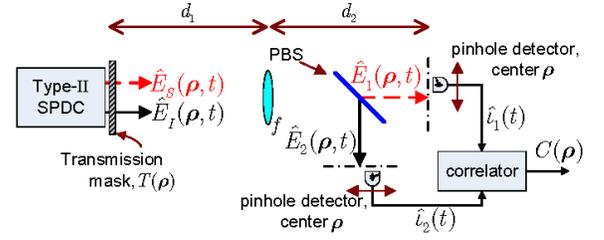}
\end{center}
\caption{(Color online) Two-photon thin-lens imaging of a transmission mask. PBS, polarizing beam splitter.} \label{TPI:NFlens}
\end{figure}

Let us now consider using an optical source with low spatial coherence and a thin lens to image a transmission mask placed at the source plane, as depicted in Fig.~\ref{TPI:NFlens}. Primary attention in our analysis of this experimental setup will be given to the resolution limitations imposed by the finite aperture of the lens, as previous work has claimed that a factor-or-two resolution improvement accrues when a broadband biphoton source is employed \cite{Shih:QI}.

As in the previous section, we shall assume an SPDC source that generates zero-mean Gaussian-state signal and idler beams whose phase-insensitive correlation spectra at the exit plane of the transmission mask are given by Eq.~\eqref{PIS:Source} and whose phase-sensitive cross-correlation spectrum at that plane is given by Eq.~\eqref{PS:Source}. The optical fields first propagate through a $d_{1}$-m-long  free-space path according to Eq.~\eqref{FS:prop}. A thin lens of radius $R$ and focal length $f$ embedded in an otherwise opaque screen is placed on this plane.  This thin lens will be assumed to be entirely free of chromatic aberration over the frequency range of interest, i.e., each frequency component of the impinging field is multiplied by $\text{circ}(|\rhovec|/R) e^{-i \omega|\rhovec|^{2}/2cf}$, where $\omega$ is a passband frequency centered around $\omega_{0}$ and the circle function is
\begin{equation}
\text{circ}(x) = \begin{cases} 1, & |x| \leq 1, \\ 0, & \text{otherwise.} \end{cases}
\end{equation}
Finally the field at the exit plane of the lens propagates $d_{2}\,$m in free space to reach the image plane, which is defined by the lens maker's formula $d_{1}^{-1} + d_{2}^{-1} = f^{-1} $. The image-plane signal and idler fields are separated by a polarizing beam splitter, after which each illuminates a pinhole photodetector located at transverse coordinate $\rhovec$ relative to its optical axis.  The resulting photocurrents are then correlated to obtain the same fourth-order field measurement given in Eqs.~\eqref{coinc} and \eqref{IntCorr} in terms of the image-plane field operators $\hat{E}_{1}(\rhovec,t)$ and $\hat{E}_{2}(\rhovec,t)$.

The overall mapping from the source-plane field operators to the image-plane field operators is linear, and therefore $\hat{E}_{1}(\rhovec,t)$ and $\hat{E}_{2}(\rhovec,t)$ are in a zero-mean jointly-Gaussian state.  With the simplifying assumption that the detectors have electrical bandwidths much broader than the source spectra \cite{BBAssumption}, we can combine and reduce Eqs.~\eqref{coinc} and \eqref{IntCorr} for the biphoton state to obtain
\begin{align}
C(\rhovec) & \!\approx\! q^{2} \eta^{2} A^{2} \langle \hat{E}_{1}^{\dagger}(\rhovec,t) \hat{E}_{2}^{\dagger}(\rhovec,t) \hat{E}_{1}(\rhovec,t) \hat{E}_{2}(\rhovec,t)  \rangle \\[5pt]
& \!=\! q^{2} \eta^{2} A^{2} \, \bigl [ K_{1,1}^{(n)}(\rhovec,\rhovec,0) K_{2,2}^{(n)}(\rhovec,\rhovec,0) \nonumber \\[5pt] &+ | K_{1,2}^{(p)}(\rhovec,\rhovec,0) |^{2}\bigr ] \,, \label{IntCorr:GaussianStates}
\end{align}
where  
\begin{align} 
K_{m,\ell}^{(n)} (\rhovec, \rhovec, \tau) & \equiv \langle \hat{E}_{m}^{\dagger}(\rhovec,t) \hat{E}_{\ell}(\rhovec,t+\tau) \rangle, \\[5pt]
K_{m,\ell}^{(p)} (\rhovec, \rhovec, \tau) &\equiv \langle \hat{E}_{m}(\rhovec,t) \hat{E}_{\ell}(\rhovec,t+\tau) \rangle, 
\end{align}
for $m,\ell = 1,2$, and Eq.~\eqref{IntCorr:GaussianStates} follows from the Gaussian moment-factoring theorem \cite{Mandel,ShapiroSun}. Furthermore, as we have determined in the previous section, for maximally-entangled Gaussian states with low-brightness and low-flux, i.e., the biphoton state, the second term in \eqref{IntCorr:GaussianStates} is much stronger than the first, permitting the approximation
\begin{equation}
C(\rhovec) \approx q^{2} \eta^{2} A^{2} | K_{1,2}^{(p)}(\rhovec,\rhovec,0) |^{2}\,. \label{TPI:NFImage}
\end{equation}
Therefore, for biphotons a photocurrent correlation measurement with broadband detectors is a means for measuring the squared magnitude of the phase-sensitive cross correlation between the image-plane field operators \cite{ShapiroSun}. In this section we demonstrate that the interesting biphoton-state results predicted for this imaging configuration are a consequence of the phase-sensitive cross correlation, and the photocurrent correlation does not play a role beyond facilitating its measurement. Hence, in the remainder of this section we shall bypass this photocurrent correlation measurement and focus directly on the phase-sensitive and phase-insensitive cross correlations between the image-plane field operators.

The frequency-domain image-plane field operators are given by the following linear transformation of the frequency-domain source-plane field operators,
\begin{equation}
\hat{\mathcal{E}}_{m}(\rhovec,\Omega) = \int\! {d}\rhovec' \, h(\rhovec,\rhovec', \omega_{0} + \Omega) \hat{\mathcal{E}}_{\ell}(\rhovec', \Omega) + \hat{\mathcal{L}}_{m}(\rhovec,\Omega)\,, \label{PSF:Freq}
\end{equation}
for $(\ell,m) = (S,1), (I,2)$, where $\hat{\mathcal{L}}_{m}(\rhovec,\Omega)$ is an auxiliary vacuum-state operator such that $\hat{E}_{m}(\rhovec,t)$ satisfies the free-field commutators given in Eqs.~\eqref{commutator1} and \eqref{commutator2}. The point-spread function $h(\rhovec,\rhovec', \omega)$, found from the Huygens-Fresnel principle and the lens transfer function, is given by 
\begin{equation}
h(\rhovec,\rhovec', \omega) = \mathcal{H}\bigl ( r(\rhovec,\rhovec'), \omega/ \omega_{0} \bigr ) e^{i\phi(\rhovec,\rhovec',\omega)}, \label{H:lens}
\end{equation}
where 
\begin{equation}
r(\rhovec,\rhovec') \equiv \frac{\omega_{0} R}{cd_{1}} | d_{1} \rhovec/d_{2} + \rhovec' |\,,  \label{r}
\end{equation}
and
\begin{equation}
\mathcal{H}(r,\xi) \equiv \frac{- \omega_{0}^{2} R^{2} \xi^{2}}{4 \pi c^{2} d_{1} d_{2}} \frac{2 J_{1}(r \xi)}{r \xi}\,, \label{H:lensAmp}
\end{equation}
with $2 J_{1}(x)/x$ for $x\geq 0$ being the well-known Airy function.  The phase term in Eq.~\eqref{H:lens} is 
\begin{equation}
\phi(\rhovec,\rhovec',\omega) \!= \omega \left (d_{1} + d_{2} + |\rhovec|^{2}/2 d_{2} + | \rhovec'|^{2}/2 d_{1} \right )/c,     \label{H:lensPhase}
\end{equation}
which incorporates the group delay arising from the $(d_{1} + d_{2})$-m propagation, and the parabolic phases at the source and image planes that are associated with the diffraction process. 

Because our concern is with the resolution limit imposed by the lens having a finite radius $R$, we will further simplify our analysis by assuming spatially-incoherent source statistics and appropriate focusing at the source plane to compensate for the parabolic phase in Eq.~\eqref{H:lensPhase}. These assumptions simplify the phase-insensitive autocorrelation functions and the phase-sensitive cross-correlation function of the $I_{0}\,$photons/m$^{2}$s signal and idler fields---given in Eqs.~\eqref{SM02:PIS} and \eqref{SM02:PS}---to
\begin{eqnarray}
S_{0}^{(n)}(\rhovec_1, \rhovec_2,\Omega)   &=& | T (\rhovec_{1} ) |^2  \bigl [ 2\pi c / (\omega_{0}+\Omega)\bigr ]^{2} I_0 \nonumber \\[5pt]  &\times& \delta(\rhovec_2-\rhovec_1) \, s^{(n)}(\Omega) \label{PISBB}
\end{eqnarray}
and 
\begin{eqnarray}
\lefteqn{S_{0}^{(p)}(\rhovec_1, \rhovec_2,\Omega) = e^{-i \omega_{0} |\rhovec_{1}|^{2}/c d_{1}} \,T^{2}(\rhovec_1 )} \nonumber  \\[5pt] &\times& \bigl [ (2\pi c)^{2} / (\omega_{0}^{2} - \Omega^{2} )\bigr ]  I_0\delta(\rhovec_2-\rhovec_1)\, s^{(p)}(\Omega)\,, \label{PSBB}
\end{eqnarray}
respectively, where $s^{(x)}(\Omega)/2\pi$, for $x=n,p$, are normalized (unity area) spectra \cite{CohArea}. 
Evaluating the phase-insensitive autocorrelations and the phase-sensitive cross correlation of the two image-plane fields at equal spatial coordinates (relative to their optical axes) and at equal times, yields
\begin{equation}
K_{m,m}^{(n)}(\rhovec,\rhovec,0)  = \int_{\mathbb{R}^{2}} \!d\rhovec' \,  |T(\rhovec')|^2 g_{n} \bigl (r(\rhovec,\rhovec') \bigr)\,,  \label{PIS:detplanecorr} 
\end{equation}
for $m=1,2$, and 
\begin{eqnarray}
K_{1,2}^{(p)}(\rhovec,\rhovec,0)  &=& e^{i \omega_{0} (2 d_{1} + 2 d_{2} + |\rhovec|^{2}/d_{2})/c} \nonumber \\[5pt] &\times& \int_{\mathbb{R}^{2}} \! d\rhovec' \, T^{2}(\rhovec')  g_{p} \bigl (r(\rhovec,\rhovec') \bigr)\,. \label{PS:detplanecorr}
\end{eqnarray}
The point-spread function in the superposition integral involving $|T(\rhovec)|^{2}$ is
\begin{eqnarray}
 g_{n}(r)  &\equiv& \frac{(2\pi c)^{2}I_0}{\omega_{0}^{2}} \int_{-\omega_{0}}^{\infty} \frac{d\Omega}{2 \pi}  s^{(n)}(\Omega)\nonumber  \\[5pt] &\times& |\mathcal{H}(r ,1+\Omega/\omega_{0})|^{2} / (1 + \Omega/\omega_{0})^{2} \,, \label{PIS:PSF}
\end{eqnarray}
and it determines the phase-insensitive autocorrelation functions. Likewise, the point-spread function in the superposition integral involving $T^{2}(\rhovec)$ is \cite{IntegralLimit}  
\begin{eqnarray}
 \lefteqn{g_{p}(r)  \equiv \frac{(2\pi c)^{2}I_0}{\omega_{0}^{2}}  \int_{-\omega_{0}}^{\omega_{0}} \frac{d\Omega}{2 \pi}  s^{(p)}(\Omega)} \nonumber \\[5pt] &\times& \mathcal{H}(r,1+\Omega/\omega_{0})  \mathcal{H}(r,1-\Omega/\omega_{0}) / (1-\Omega^{2}/\omega_{0}^{2}),\, \label{PS:PSF} 
\end{eqnarray}
and it controls the phase-sensitive cross-correlation function. Therefore, apart from an unimportant parabolic phase factor, the most important difference between phase-insensitive and phase-sensitive coherence propagation---insofar as two-photon thin-lens imaging is concerned---is the frequency coupling between $\pm \Omega/\omega_{0}$ that is present in Eq.~\eqref{PS:PSF} but is absent from Eq.~\eqref{PIS:PSF}.  In the quasimonochromatic limit, however, this coupling becomes insignificant, because 
\begin{equation} 
1 \pm \Omega/\omega_{0} \approx 1\,,
\end{equation}
so that  $g_{n}(r) = g_{p}(r)$ prevails whenever $s^{(n)}(\Omega) = s^{(p)}(\Omega) $. Thus the quasimonochromatic point-spread function for the phase-insensitive correlation is identical to its quasimonochromatic phase-sensitive counterpart.

\begin{figure}[t]
\centering
\subfigure[Main-lobe behavior.]{
		\includegraphics[width=2.8in]{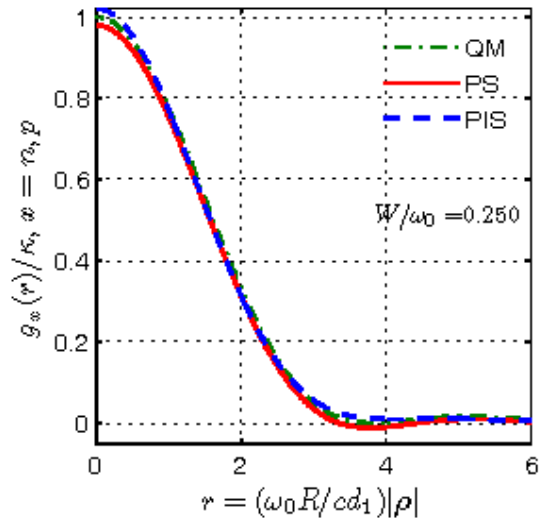}
		\label{PSNFKernel:1} 
	}\\
\subfigure[Tail behavior (logarithmic scale).]{
		\includegraphics[width=2.8in]{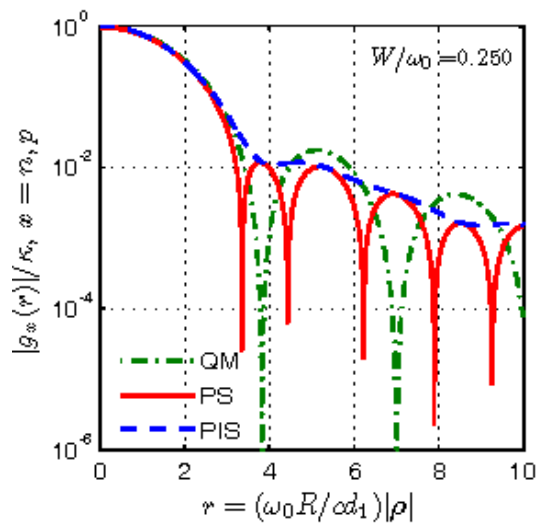}
		\label{PSNFKernelLog:1}
	}
	\caption{(Color online) Comparison of the imaging point-spread functions for phase-insensitive (PIS) and phase-sensitive (PS) correlations when $W/\omega_{0}=0.25$, and when the imaging source is quasimonochromatic (QM). The normalizing coefficient is $\kappa \equiv I_{0} \omega_{0}^{2} R^{4} / 4 c^{2} d_{1}^{2} d_{2}^{2}$.}
	\label{Plot:BB1}
\end{figure}

\begin{figure}[ht]
\centering
\subfigure[Main-lobe behavior.]{
		\includegraphics[width=2.8in]{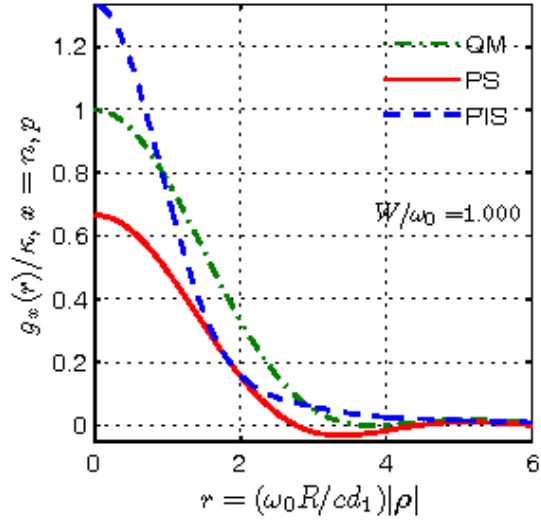}
		\label{PSNFKernel:4}
	} \\
\subfigure[Tail behavior (logarithmic scale).]{
		\includegraphics[width=2.8in]{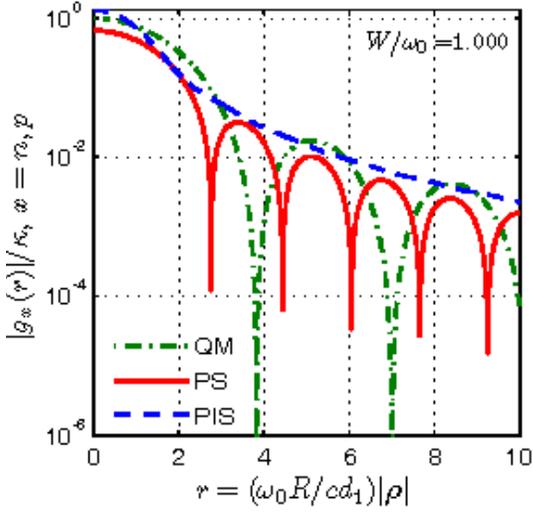}
		\label{PSNFKernelLog:4}
	}
	\caption{(Color online) Comparison of the imaging point-spread functions for phase-insensitive (PIS) and phase-sensitive (PS) correlations at the ultimate theoretical limit of SPDC bandwidth ($W/\omega_{0} = 1$), and the quasimonochromatic (QM) limit. The normalizing coefficient is $\kappa \equiv I_{0} \omega_{0}^{2} R^{4}/4 c^{2} d_{1}^{2} d_{2}^{2}$.}  
	\label{Plot:BB2}
\end{figure}

However, $g_{n}(r)$ and $g_{p}(r)$ begin to differ as the bandwidth of the source increases. Suppose that the normalized phase-insensitive and phase-sensitive source spectra are both taken to be flat over a $2W$-bandwidth window,  i.e., 
\begin{equation}
s^{(n)}(\Omega) = s^{(p)}(\Omega) = \begin{cases} \pi/W, & |\Omega| < W, \\ 0, & \text{otherwise.} \end{cases}
\end{equation} 
Substituting this expression into Eqs.~\eqref{PIS:PSF} and \eqref{PS:PSF} permits us to express the point-spread functions as the following  dimensionless integrals,
\begin{equation}
 g_{n}(r)  = \frac{I_{0} \omega_{0}^{3} R^{4} }{2 c^{2} d_{1}^{2} d_{2}^{2} W} \int_{-W/\omega_{0}}^{W/\omega_{0}} {d}u \,  \frac{J_{1}^{2} \bigl(r (1+u) \bigr)  }{r^{2}} \,, \label{PIS:nodimen}
\end{equation}
and
\begin{eqnarray}
 g_{p}(r)  &=& \frac{I_{0} \omega_{0}^{3} R^{4} }{2 c^{2} d_{1}^{2} d_{2}^{2} W} \int_{-W/\omega_{0}}^{W/\omega_{0}} {d}u \nonumber \\[5pt] &\times& \frac{J_{1}\bigl (r (1+u) \bigr ) }{r } \frac{J_{1}\bigl (r (1-u) \bigr ) }{r }\,.   \label{PS:nodimen}
\end{eqnarray}
In the quasimonochromatic limit, in which $W/\omega_{0} \ll 1$ holds, both point-spread functions simplify to 
\begin{equation}
g_{n}(r) = g_{p}(r) = \frac{I_{0} \omega_{0}^{2} R^{4} }{4 c^{2} d_{1}^{2} d_{2}^{2}} \Bigl ( \frac{2 J_{1}(r)}{r} \Bigr )^{2}\,. \label{QM:PSF}
\end{equation} 
Hence with a quasimonochromatic source there is no difference between the image of a real-valued transmission mask acquired with phase-insensitive (thermal) illumination or phase-sensitive (classical or quantum) illumination.

The point-spread functions for broader bandwidth sources are plotted in Figs.~\ref{Plot:BB1} and \ref{Plot:BB2} at two different $W$ values. The $W= \omega_{0}/4$ phase-sensitive point-spread function, shown in Fig.~\ref{Plot:BB1}, represents imaging performance for unusually broadband SPDC \cite{ODonnell}.  The $W=\omega_{0}$ phase-sensitive point-spread function, plotted in Fig.~\ref{Plot:BB2}, represents imaging performance at the ultimate theoretical limit of SPDC bandwidth.  The point-spread functions in these figures show that the peak amplitude of the phase-insensitive function increases to
\begin{equation}
g_n(0) = \frac{I_{0} \omega_{0}^{2} R^{4} }{4 c^{2} d_{1}^{2} d_{2}^{2}} \left ( 1 + \frac{W^{2}}{3 \omega_{0}^{2}} \right )\,,
\end{equation}
whereas that of the phase-sensitive point-spread function attenuates to 
\begin{equation}
g_p(0) = \frac{I_{0} \omega_{0}^{2} R^{4} }{4 c^{2} d_{1}^{2} d_{2}^{2}} \left ( 1 - \frac{W^{2}}{3 \omega_{0}^{2}} \right )\,,
\end{equation}
relative to the peak amplitude in the quasimonochromatic limit as the source bandwidth increases.  The $(1+u)^{2}$ factor multiplying the frequency-resolved Airy patterns in Eq.~\eqref{PIS:nodimen}, where $|u|<W/\omega_{0}\leq 1$, is responsible for the increase in $g_n(0)$ with increasing source bandwidth. This scaling favors the {\em blue}-detuned frequency contributions to the phase-insensitive point-spread function.  So, because of the quadratic scaling, the average of the amplitude increase for $u>0$ versus the amplitude decrease for $u<0$ is greater than one. Thus, the peak value of $g_n(0)$ increases with increasing source bandwidth. On the other hand, the scaling for the frequency-resolved Airy patterns in the integrand of Eq.~\eqref{PS:nodimen} is  $(1-u)(1+u) = 1-u^{2}$, implying that  {\em all} detuned frequencies are attenuated, which results in $g_p(0)$ decreasing as the source bandwidth increases.

Figures~\ref{Plot:BB1}(a) and \ref{Plot:BB2}(a) indicate that some narrowing of the main lobe occurs for both the broadband phase-sensitive and phase-insensitive point-spread functions, with the latter suffering from a slower-decaying tail.   From a practical perspective this main-lobe narrowing behavior is of little interest.  Taking the resolution to be set by the first zero in the point-spread function, Fig.~\ref{Plot:BB1}(b) shows that the resolution benefit offered by broadband phase-sensitive as compared to quasimonochromatic (phase-sensitive or phase-insensitive) imaging is merely a factor of $1.14$.   From this figure we also note that the tail of the phase-insensitive point-spread function traces the envelope of the oscillations in the phase-sensitive point-spread function, so no appreciable loss of resolution results from the former's slowly-decaying tail.  Even at the  $W = \omega_{0}$ ultimate bandwidth limit, the resolution improvement offered by the broadband phase-sensitive point-spread function---as found from Fig.~\ref{Plot:BB2}(b)---is only a factor of $1.38$.  In this limit, the tail of the broadband phase-insensitive point-spread function falls off somewhat---but not dramatically---slower than the envelope of the phase-sensitive point-spread function's oscillations.  

\begin{figure}[t]
\begin{center}
\includegraphics[width= 3.0 in]{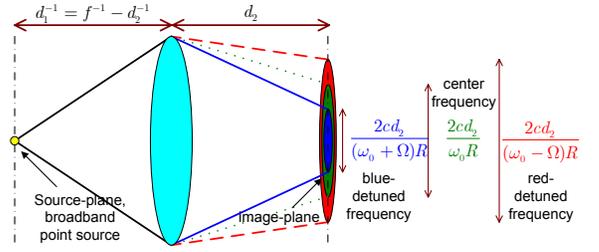}
\end{center}
\caption{(Color online) Image-plane spot diameters for different frequency components of a broadband point source.} \label{TPI:LensDiffFreq}
\end{figure}

The difference between the behavior of the phase-insensitive and phase-sensitive point-spread functions as a function of the source bandwidth deserves closer examination to understand the underlying physics. Recall that a source generating a complex-stationary baseband field around a center frequency $\omega_{0}$ is a superposition of monochromatic field components which have phase-insensitive autocorrelations at each frequency and phase-sensitive cross correlations between frequencies that sum to $2 \omega_{0}$. Thus, the phase-insensitive correlation, measured at a given spatiotemporal coordinate $(\rhovec,t)$, is a superposition of all the different autocorrelations at detunings $\Omega$ over the fluorescence bandwidth of the source. On the other hand, the phase-sensitive correlation measured at $(\rhovec,t)$ is a superposition of all the {\em cross correlations} between frequency components detuned by $\pm \Omega$, over the phase-sensitive bandwidth of the source.  Now, consider a point source at the source-plane that emits signal and idler fields that have nonzero phase-insensitive autocorrelations and phase-sensitive cross correlation.  From Eq.~\eqref{PSF:Freq}, signal and idler frequency components at $\omega = \omega_{0} + \Omega$ will yield image-plane spots  with common radius $c d_{2}/(\omega_{0} +\Omega) R$. Thus, as shown in Fig.~\ref{TPI:LensDiffFreq}, lower frequency components produce broader spots on the image plane than do higher frequency components. The phase-insensitive autocorrelation---of either image-plane field---measured by scanning a point detector on the transverse plane, therefore decays slowly as $|\rhovec|$ increases, because of the large spots from the lower frequencies. 
However, this slowly-decaying tail does not cause a significant increase in the point-spread function's width, because the quadratic weighting coefficient in Eq.~\eqref{PIS:nodimen} accentuates blue-detuned frequencies and attenuates those that are red detuned. On the other hand, if we are measuring the phase-sensitive cross correlation between the signal and idler fields \cite{AutovsCross:2}, we are in effect measuring the superposition of the cross correlations between the $\omega_{0}+ \Omega $ signal-field component and the $\omega_{0}- \Omega $ idler-field component, where $\Omega \in [-\omega_{0}, \omega_{0}]$. For $\Omega>0$, the former yields a narrow spot of radius $c d_{2}/(\omega_{0} +\Omega) R$, and the latter yields a broad spot of radius $c d_{2}/(\omega_{0} -\Omega) R$. Because the phase-sensitive cross correlation is given by their product, however, the narrower radius from the higher frequency determines the radius within which there is appreciable phase-sensitive coherence. Furthermore, this coherence radius is symmetric in $\Omega$, so, as the phase-sensitive bandwidth of the source increases, the width of the image-plane phase-sensitive point-spread function decreases. However, the weighting coefficient $1-u^{2}$ in Eq.~\eqref{PS:nodimen} counteracts this advantage by attenuating the frequency contributions with higher detunings, such that the net reduction in the main lobe's width is very small.

Notice that we have made no reference to the classical or quantum nature of the source in explaining the physics governing the point-spread functions' width. Thus, this effect is entirely a consequence of phase-sensitive versus phase-insensitive source correlations and scalar paraxial diffraction theory, both of which are valid in the classical and quantum theories of light. The quantum nature of the fields, therefore, does not play a role in determining the resolution capabilities of thin-lens correlation imaging, regardless of whether the source has phase-sensitive or phase-insensitive coherence. 
However, for particular measurement schemes, nonclassical field states may offer contrast advantages akin to those found in the previous section for diffraction-pattern imaging.  In particular, in this section we have determined that the phase-sensitive correlation differs from its phase-insensitive counterpart only in the broadband limit. Thus, if we opt to utilize a photocurrent correlation measurement, then the contrast will be significantly better when the broadband fields' state is maximally-entangled (nonclassical) and has low-brightness, which encompasses the biphoton state. 

\section{Discussion}
\label{CH6:discussion}

SPDC with vacuum-state inputs generates signal and idler fields in a zero-mean jointly-Gaussian state, with nonzero phase-insensitive autocorrelations and a phase-sensitive cross correlation that fully determine their joint state. When the output state is driven to the low-brightness, low-flux limit, this Gaussian state becomes equivalent to a dominant vacuum state plus a weak biphoton contribution, in which the biphoton wave function equals the phase-sensitive cross correlation between the signal and idler fields. 
On the other hand, classical imagers have traditionally utilized optical sources in thermal states or coherent states, both of which are Gaussian states but have only nonzero phase-insensitive correlations. Hence, quantum imaging experiments that rely on biphoton sources, as well as conventional classical imaging configurations, can be unified and generalized by studying the imaging characteristics of Gaussian-state sources.
Furthermore, such states are fully characterized by their first and second moments, and are closed under linear transformations on the field operators. So, imaging configurations utilizing Gaussian-state sources, linear optical elements and free-space propagation can be fully understood in both the classical and quantum regimes by tracking the evolution of the first and second moments of the fields from the source plane to the detection planes.

A particularly relevant distinction that has been overlooked in most previous work is the phase-sensitive nature of the correlation between the two photons in a biphoton state, as opposed to the phase-insensitive correlation that is present between thermal-state fields. Phase-sensitive coherence has propagation characteristics that differ from those of phase-insensitive coherence. Furthermore, complex-stationary phase-sensitive correlations have cross-frequency couplings that are not present in complex-stationary phase-insensitive correlations. Distinctions such as these often underlie the interesting observations and theoretical predictions in quantum imaging. However, phase-sensitive coherence is not exclusive to nonclassical states (such as the biphoton). Classical Gaussian states (random mixtures of coherent states) may very well have nonzero phase-sensitive  correlations, and those features in quantum imaging that stem from the phase-sensitive coherence between the two photons in a biphoton state can be replicated with classical phase-sensitive sources, as we have previously demonstrated for optical coherence tomography and ghost imaging \cite{ErkmenShapiro:PCOCT,ErkmenShapiro:GhostImaging2}.

In this paper we continued to distinguish the truly quantum phenomena in quantum imaging theory and experiments from the phase-sensitive coherence phenomena that can be exploited both in the classical and quantum regimes. Toward this end, we performed Gaussian-state analyses of two significant experimental configurations in biphoton-state quantum imaging. In Sec.~\ref{CH6:FarFieldDiffraction} we showed that the factor-of-two spatial compression in the far-field diffraction pattern  of a transmission mask placed at the source plane is precisely due to the phase-sensitive cross correlation between the signal and idler fields, in {\em both} the classical and quantum regimes.  Indeed, the only significant difference---insofar as this experiment is concerned---between phase-sensitive classical and quantum sources is the image contrast when photocurrent correlation measurements are employed. Narrowband classical Gaussian states can achieve acceptable contrast, but the contrast degrades severely when the source is broadband. On the other hand, with low-brightness quantum Gaussian states that are maximally-entangled, the contrast is high for both narrowband and broadband sources. 
Note that the strength of the background in the signature may be a relevant factor in determining whether a classical or quantum source is more desirable for a particular application. For example, in photolithographic applications, in which extraneous noise may be eliminated by virtue of operation in a controlled environment, a biphoton-state source in combination with a two-photon absorber at the detection plane generates an optical image with no background, whereas a classical phase-sensitive source yields significant background that requires postdetection processing prior to etching. Hence the contrast advantage offered by the biphoton state---which cannot be replicated by classical phase-sensitive light---is a desirable feature in this case.

In Sec.~\ref{CH6:BroadbandImaging} we compared thin-lens imaging of a source plane transmission mask using incoherent phase-insensitive light to the same imaging arrangement using phase-sensitive light. When the sources are narrowband (quasimonochromatic), the point-spread functions of the two cases turn out to be identical, yielding no resolution difference between the various source possibilities. As the source bandwidth increases, the point-spread functions for the phase-insensitive and phase-sensitive correlation functions become narrower, with the phase-insensitive point-spread function developing a more slowly decaying tail. The differences between the two cases stem from complex-stationary source statistics, and frequency-dependent free-space diffraction. Once again, the biphoton state facilitates a high-contrast image and a convenient measurement apparatus (coincidence counting) for detecting phase-sensitive correlation, but it is not responsible for the physics governing the changes to the point-spread functions. 

Although Sec.~\ref{CH6:FarFieldDiffraction} concentrated on a biphoton proof-of-principle experiment for quantum lithography, the driving motivation for quantum optical lithography is the $N$-fold improvement in etching resolution that is predicted for a system using $N00N$ states \cite{Boto}.  The $N00N$ state is an equal-weight superposition of two pure states:  an $N$-photon signal field and a vacuum-state idler, plus a vacuum-state signal field and an $N$-photon idler. The $N00N$ state is  nonclassical; its $P$ representation in terms of coherent states is not a proper probability density.  The $N=2$ case can be achieved with biphoton states, viz., the output of a 50-50 beam splitter when the two inputs are the signal and idler fields from SPDC operating in the low-brightness, low-flux regime. Unfortunately, generating $N00N$ states for $N>2$ has proven challenging.   Thus far the interference fringes for the $N=3$ and $N=4$ cases have been demonstrated in proof-of-principle experiments \cite{Mitchell:3003state,Walther:4004state}, showing factors of $3$ and $4$ fringe compression respectively, and efforts to generate higher orders continue.  $N00N$ states with $N>2$ are not Gaussian states or any limiting form of Gaussian states, because their second-order moments do not determine the state. Therefore, the Gaussian-state analysis presented in this paper does not generalize to $N00N$ states with $N >2$. As a result, to better appreciate the fundamental physics that leads to improved resolution with these sources, it is of great interest to develop a unifying coherence theory for higher-order moments of continuous field operators, and perform an analysis for these moments to determine whether the advantages observed with these states are truly due to their nonclassical nature or due to a measurement of a $2N$th-order moment of the field operator.

The analyses presented in Secs.~\ref{CH6:FarFieldDiffraction} and \ref{CH6:BroadbandImaging} reveal that the physics governing the resolution improvement in Fourier-plane and thin-lens imaging are {\em different}.  Specifically, the improvement in resolution that is observed in the Fourier-plane measurement is due to the difference in paraxial propagation of phase-sensitive and phase-insensitive correlations, and is valid in the quasimonochromatic regime as well as in the broadband regime. Furthermore, to observe this effect with classical fields, it is preferable to utilize narrowband sources. On the other hand, the marginal improvement in resolution observed in Sec.~\ref{CH6:BroadbandImaging} is a {\em strictly} broadband effect that manifests itself in complex-stationary phase-sensitive and phase-insensitive correlation functions. The two experiments capitalize on different properties of phase-sensitive coherence and therefore they are not experiments demonstrating equivalent physical principles.

In conclusion, we have presented a unified Gaussian-state analysis of two transverse imaging configurations, one that images the far-field diffraction pattern of a source-plane transmission mask, and one that performs thin-lens imaging of that source-plane transmission mask. 
We have shown that the far-field diffraction patterns obtained with classical phase-sensitive Gaussian-state light and nonclassical Gaussian-state light with low-brightness---such as the biphoton---differ only in contrast, viz., the fringe compression is a classical phenomenon owing to the far-field diffraction of phase-sensitive coherence.  In the second experiment, we have demonstrated that the cross-frequency coupling in complex-stationary broadband phase-sensitive light---whether classical or quantum---leads to a slightly narrower point-spread function than that obtained with quasimonochromatic phase-sensitive or phase-insensitive light. However, because of the enormous bandwidth that is necessary to observe any appreciable change in the point-spread function, contrary to what is stated in \cite{Shih:QI}, there is no practical advantage  to be gained from broadband operation in this image acquisition configuration.

\section*{Acknowledgement}
This work was supported by the U. S. Army Research Office MURI Grant W911NF-05-1-0197.


\begin{thebibliography}{19}
\bibitem{classicalstates}In this regard we emphasize that we follow the quantum optics convention in which a field state is deemed classical if it has a proper $P$ representation, i.e., if it is a coherent state or a classically-random mixture of such states.  As detailed in \cite{ErkmenShapiro:GhostImaging2}, this definition is appropriate for our imaging work because such states lead to quantum photodetection
statistics that coincide with those obtained from semiclassical photodetection, in which the light field is treated as a (possibly random) classical field and the discreteness of the electron charge leads to shot noise in the detection process.  Field states that do not possess proper $P$ representations are then deemed nonclassical, and, in general, require that the quantum theory of photodetection be employed to  derive their measurement statistics.
\bibitem{coherence_separability} The signal and idler fields obtained from spontaneous parametric downconversion (SPDC) will not have space-time correlation functions that take these specific forms, because of the space-time coupling that is inherent in SPDC phase-matching relations.  However, these assumptions, which are commonly-employed in coherence theory, simplify the analytical treatment without compromising the fundamental physics.
\bibitem{ApproxValidation}  Our primary interest in this paper is the phase term in the Huygens-Fresnel principle (rather than the frequency dependence of the leading coefficient) over the frequency range of the source correlation spectrum, which is what motivates this approximation. More care will be paid to the leading coefficient when we treat broadband imaging within the Fig.~\ref{TPI:NFlens} configuration.
\bibitem{QuantLithFootnote}Type-I phase matching, rather than type-II phase matching, would be used in SPDC-based quantum lithography.  The resulting co-polarized signal and idler photons would then illuminate a two-photon absorbing photoresist.  In Fig.~1, type-II phase matching permits the use of a polarizing beam splitter and photon coincidence counting to stand in for the two-photon absorbing photoresist.
\bibitem{AutovsCross} Because both the signal and idler beams undergo identical transformations, the autocorrelation results reported in Sec.~\ref{CH6:CoherenceTheory} apply to cross correlation propagation as well.
\bibitem{NearvsFar:Coherent} Quasimonochromatic, paraxial diffraction of a coherent beam with center wavenumber $k_{0}$ and source-plane intensity radius $a_{0}$ has near- and far-field regimes that are distinguished  by a single Fresnel number $D_{C} = k_{0} a_{0}^{2}/2 L$ \cite{Mandel}. The far-field regime, given by $D_{C} \ll 1$, coincides with that for low-coherence phase-sensitive correlation diffraction. However, the near-field condition for a coherent beam, $D_{C} \gg 1$, is much less stringent than that for low-coherence phase-sensitive correlation diffraction, which corresponds to $k_{0} \rho_{0}^{2}/2 L \gg 1$, where $\rho_{0}$ is the coherence length of the source.
\bibitem{LowFluxCond} Because pinhole detectors are utilized at the detection plane, the Fresnel number product between the source-plane pupil and the photodetector pupil is much less than unity, i.e., $D_{f} = (k_{0} a_{0}^{2} / 2 L) (k_{0} A / 2\pi L) \ll 1 $, so only a single spatial mode couples appreciably to the photodetectors at each $\rhovec$ \cite{Slepian:Apodization}. Hence, the low-flux condition stated here should be interpreted as low-flux per spatial-mode of the source. In particular, very high contrast is achieved even when multiple photon-pairs are generated at the source, as long as at most one photon-pair occupies each spatial mode.
\bibitem{BBAssumption} Although this broadband detector assumption will certainly be valid for narrowband sources, it will almost surely fail for the very broadband sources to be considered later. Nevertheless, it is a convenient way to focus our attention on the essential physics, viz., imaging via $|K^{(p)}_{1,2}(\rhovec,\rhovec,0)|^{2}$ or $|K^{(n)}_{1,2}(\rhovec,\rhovec,0)|^{2}$.
\end{thebibliography}
\end{document}